\documentclass[aps,10pt,twocolumn,showpacs,preprintnumbers,amsmath,amssymb,prb]{revtex4-1}

\usepackage{color}
\usepackage{graphicx}
\usepackage{txfonts}
\usepackage[colorlinks,citecolor=blue,linkcolor=blue,urlcolor=blue,plainpages=false,pdfpagelabels]{hyperref}

\newcommand{\ket}[1]{|#1 \rangle}
\newcommand{\eso}{\epsilon_{\text{SO}}}
\newcommand{\hc}{\text{h.c.}}
\newcommand{\lso}{\ell_{\rm so}}
\newcommand{\Bcrit}{B_{\rm crit}}

\newcommand{\tH}{\tilde{H}}
\newcommand{\oy}{z'}
\newcommand{\vS}{\vec{S}}
\renewcommand{\vr}{\vec{r}}

\begin{document}

\title{Helical gaps in interacting Rashba wires at low electron densities}

\author{Thomas~L.~Schmidt}
\author{Christopher~J.~Pedder}
\affiliation{Physics and Materials Science Research Unit, University of Luxembourg, L-1511 Luxembourg}

\date{\today}

\begin{abstract}
Rashba spin-orbit coupling and a magnetic field perpendicular to the Rashba axis have been predicted to open a partial gap (``helical gap'') in the energy spectrum of noninteracting or weakly interacting one-dimensional quantum wires. By comparing kinetic energy and Coulomb energy we show that this gap opening typically occurs at low electron densities where the Coulomb energy dominates. To address this strongly correlated limit, we investigate Rashba wires using Wigner crystal theory. We find that the helical gap exists even in the limit of strong interactions but its dependence on electron density differs significantly from the weakly interacting case. In particular, we find that the critical magnetic field for opening the gap becomes an oscillatory function of electron density. This changes strongly the expected signature of the helical gap in conductance measurements.
\end{abstract}


\maketitle

\section{Introduction}

The past years have brought a rapid growth of interest in quantum wires with Rashba spin-orbit coupling (RSOC). Much of this activity results from the discovery that, if subjected to the proximity effect of a nearby superconductor and a magnetic field, such wires can host Majorana bound states at their ends.\cite{oreg10,lutchyn10}
Experimental signatures of these elusive quantum states have already been found in indium arsenide (InAs) or indium antimonide (InSb) quantum wires.\cite{mourik12,deng12,das12,rokhinson12,albrecht16}

While many of the expected properties of Majorana bound states have been verified, ruling out all possible alternative explanations still requires a better understanding of the wires used in experiments. Therefore, more experimental effort has recently been devoted to the investigation of normal-conducting Rashba wires and in particular to the characterization of their RSOC itself.\cite{vanweperen15,montemurro15,kammhuber16,zhang16} A rather straightforward experimental signature of RSOC would be a ``helical gap'', i.e., the opening of a partial gap in the energy spectrum of a Rashba wire in response to an applied magnetic field perpendicular to the Rashba axis, see Fig.~\ref{fig:Spectrum}. Indications of such a gap have already been found in another material,\cite{quay10} and experimental efforts in InAs and InSb quantum wires are currently underway.

In its simplest form, the helical gap can be understood based on a single-particle theory. It is evident, however, that this gap appears near the band bottom and thus at low electron densities $\rho \approx (\pi \lso)^{-1}$, where $\lso$ is the spin-orbit length. Not only does this present a major challenge for experimentalists, it also renders the theoretical description in the presence of electron-electron interactions more complicated. A direct comparison shows that at the required electron densities the Coulomb energy actually exceeds the kinetic energy of the electrons. In this case, the energy range accessible to Luttinger liquid (LL) theory is exponentially suppressed as a function of density.\cite{meyer09,imambekov12} For electrons without spin-orbit coupling, this limit was reviewed in detail recently in the context of spin-incoherent LLs.\cite{fiete07}

The low-density limit mandates a theoretical description in terms of a 1D Wigner crystal.\cite{wigner34,glazman92,schulz93} This approach has advanced considerably over the past decade,\cite{matveev04,matveev04_2,matveev07,meyer09,meng11,ziani13,ziani13b,yakimenko13} and experiments have already shown signs of Wigner crystal phases in quantum wires,\cite{goni91,hew09,yamamoto12,yamamoto15} and carbon nanotubes.\cite{deshpande08}

To study the helical gap, we extend the theory of 1D Wigner crystals to systems with RSOC. We start with a short discussion of the noninteracting case, followed by an estimate of the Coulomb energy. Next, we derive the effective Hamiltonian governing the charge and spin sectors of the Rashba wire at low densities. We find that the spin Hamiltonian has a spectral gap for magnetic fields above a critical field $\Bcrit(\rho)$ which depends in a nontrivial way on the electron density $\rho$. Finally, we calculate the differential conductance of the interacting quantum wire which is the most accessible experimental probe of the helical gap.

\section{Model}

\begin{figure}[t]
  \centering
  \includegraphics[width=\columnwidth]{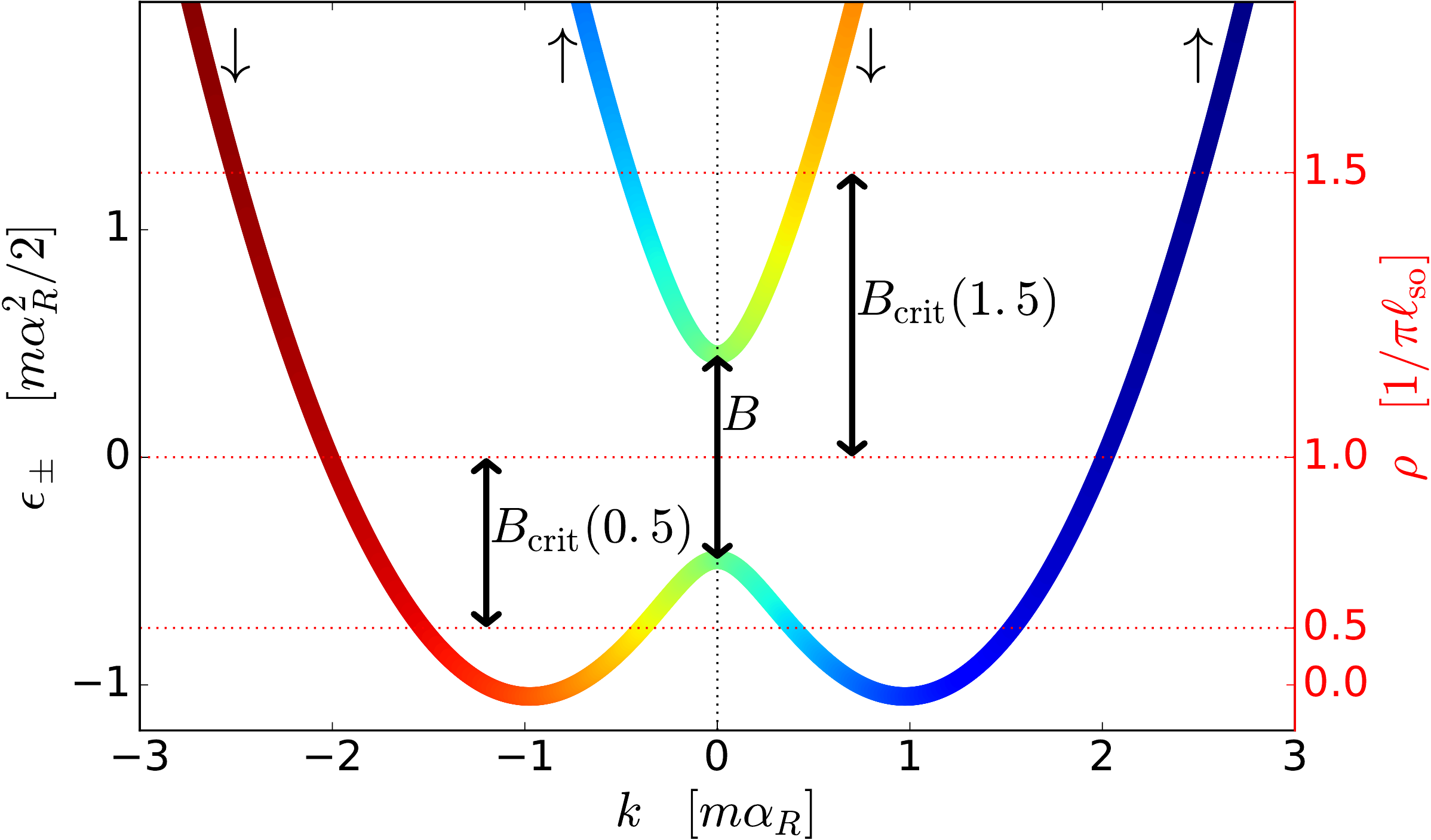}
  \caption{Single-particle spectra $\epsilon_\pm(k)$ for a weak magnetic field ($g \mu_B B < m \alpha_R^2$). The color coding shows the spin orientation as a function of momentum. The corresponding density axis is shown on the right, and values of $\Bcrit(\rho)$ for two densities are indicated.}
  \label{fig:Spectrum}
\end{figure}

We start by considering a single electron with band mass $m$ moving in a one-dimensional wire along the $z$ direction. In the presence of RSOC with strength $\alpha_R$, and a magnetic field perpendicular to the wire in the $x$ direction, the single-particle Hamiltonian and its spectrum read\cite{oreg10,lutchyn10} (using $\hbar = 1$)
\begin{align}
    H_1 &= \frac{p^2}{2m} - \alpha_R p \sigma^z - g \mu_B \vec{B} \cdot \vec{S}, \label{eq:H} \\
    \epsilon_\pm(k) &= \frac{k^2}{2m} \pm \sqrt{(g \mu_B B/2)^2 + \alpha_R^2 k^2}, \label{eq:spec}
\end{align}
where $p$ is the momentum operator, and the electron spin is given by $\vec{S} = \vec{\sigma}/2$ where $\vec{\sigma} = (\sigma^{x}, \sigma^{y}, \sigma^{z})$ is the vector of Pauli matrices. The magnetic field $\vec{B} = (B, 0, 0)$, where we assume $B > 0$, gives rise to the Zeeman energy $g \mu_B B$ which depends on the $g$ factor and the Bohr magneton $\mu_B$. The appearance of the helical gap is an immediate consequence of the spectrum (\ref{eq:spec}), which is shown in Fig.~\ref{fig:Spectrum}. For small magnetic fields ($g \mu_B B < m \alpha_R^2$) the spectrum develops a local maximum and a gap of width $g \mu_B B$ at $k = 0$, whereas the outer modes remain largely unaffected.

To connect to later results for the interacting case, we rephrase the condition for a helical gap in terms of the electron density $\rho$. At zero magnetic field, the spectrum consists of two shifted parabolas and the chemical potential can be written as a function of the electron density as $\mu(\rho) = (\pi \rho)^2/(8m) - m \alpha_R^2/2$. We define the critical field $\Bcrit$ as the minimum magnetic field needed to gap out the modes at a given chemical potential $\mu$. Hence, we find $g \mu_B \Bcrit = |\mu|$, which expressed in terms of electron density reads
\begin{align}\label{eq:BcritNI}
    g \mu_B \Bcrit(\rho) = E_F \left| 1 - \left(\frac{\varphi}{\pi}\right)^2\right|, \qquad
    \varphi = \frac{1}{\rho \lso},
\end{align}
where we defined the Fermi energy $E_F = (\pi \rho)^2/(8m)$ and the spin-orbit length $\lso = (2 m \alpha_R)^{-1}$. Therefore, at the critical density $\rho = (\pi\lso)^{-1}$ (corresponding to $\mu=0$), an infinitesimal magnetic field opens the helical gap. Away from this density, a finite magnetic field $\Bcrit \propto |\varphi - \pi|$ is needed. The size of the gap as a function of the deviation from the critical field, $\delta B = B - \Bcrit$ is given by,
\begin{align}\label{eq:Delta}
    \Delta(\rho, B) = g \mu_B \delta B
\end{align}
The simplest experimental signature of the helical gap is a dip in the zero bias-conductance as a function of electron density. At zero temperature, it is given by
\begin{align}\label{eq:G}
    G(\rho, B) = 2 G_0 - G_0 \Theta\left[B - \Bcrit(\rho)\right]
\end{align}
where $\Theta(x)$ denotes the Heaviside function and $G_0 = e^2/h$ is the conductance quantum. At a given electron density, the dip in the conductance remains visible up to temperatures $T \approx \Delta(\rho,B)$.\cite{schmidt13b} The case of weak interactions can be approached using bosonization\cite{braunecker09,braunecker09b,braunecker10,gangadharaiah11,stoudenmire11,gambetta15} which predicts a renormalization of system parameters but does not change the structure of the helical gap qualitatively. A comparison of our results to bosonization results is shown in App.~\ref{sec:Bosonization}.

Experimental estimates for the spin-orbit lengths are in the range of $\lso \approx 200 \text{nm}$,\cite{hernandez10,mourik12} so observing the helical gap requires rather low densities $\rho \approx (\pi\lso)^{-1}$. Such low electron densities \emph{increase} the effect of the Coulomb potential $V(z) = e^2/(\epsilon |z|)$, where $\epsilon$ is the dielectric constant and $e$ the electron charge. This is a peculiar consequence of Fermi statistics, which entails that the kinetic energy per particle scales as $E_{\rm kin} \propto \rho^2$, while the Coulomb energy per particle scales as $E_{\rm pot} \propto \rho$. More precisely, the Coulomb energy dominates for densities $\rho a_B \ll 1$, where $a_B = 4 \pi \epsilon/(m e^2)$ is the Bohr radius. The bare Coulomb repulsion is usually screened at large distances. If screening is due to a gate at a distance $d$ from the wire, the potential reads $V(z) = (e^2/\epsilon)( |z|^{-1} - |z^2 + 4d^2|^{-1/2} )$ and the Coulomb energy dominates if\cite{matveev04,matveev04_2,matveev07,meyer09,meng11}
\begin{align}\label{eq:rho_range}
    a_B \ll \frac{1}{\rho} \ll \frac{d^2}{a_B}.
\end{align}
Equation~(\ref{eq:rho_range}) specifies the density range where the results we derive below are applicable. For InSb ($\epsilon \approx 17$, $m \approx 0.015 m_{e}$, see Ref.~[\onlinecite{mourik12}]), one finds $a_B \approx 60 \text{nm}$, for InAs ($\epsilon \approx 15$, $m \approx 0.033 m_e$, see Ref.~[\onlinecite{hernandez10}]), the Bohr radius is $a_B \approx 25\text{nm}$. Screening is discussed in more detail in App.~\ref{sec:Screening}, and we show there that metallic gates comparable in size to the nanowire are insufficient for screening. The most important contribution to screening is thus provided by a macroscopic gate. Wigner crystal formation in carbon nanotubes has been observed for a gate distance on the order of $d \approx 600\text{nm}$.\cite{deshpande08} A similar gate distance in a Rashba nanowire would lead to $d^2/a_B \approx 10^{4}\text{nm}$. Hence, near the critical electron density $\rho^{-1} \approx \pi\lso \approx 600 \text{nm}$ required for the observation of the helical gap, the inequality (\ref{eq:rho_range}) is fulfilled and the Coulomb repulsion indeed dominates over the kinetic energy. We will therefore develop a theoretical model of the helical gap taking into account the strong effect of the Coulomb repulsion.

We would like to point out that most investigations on Rashba wires have so far focused on Majorana wires, where a nearby superconductor screens the Coulomb interaction. In that case, the interactions can be modelled using bosonization and have much weaker effects. In contrast, this article focuses on bare wires where screening is less efficient. In this case, the fact that $d \gg a_B$ opens a large density window (\ref{eq:rho_range}) for the Wigner crystal formation.

\section{Wigner crystal theory}

To develop a 1D Wigner crystal theory for systems with RSOC, we start by considering a system of $N$ electrons, each of which is described by the Hamiltonian (\ref{eq:H}), and add the interaction term $V(z_m - z_n)$, where $z_n$ denotes the position of the $n$th particle. Moreover, it is convenient to perform a unitary transformation $U = \prod_n \exp\left(2 i m \alpha_R z_n S^z_n\right)$ on Eq.~(\ref{eq:H}) to gauge away the Rashba term at the expense of turning the constant magnetic field into a spiral magnetic field in the spin-$x-y$ plane. Importantly, this transformation commutes with the interaction Hamiltonian. Hence, the transformed Hamiltonian reads (see App.~\ref{sec:Ham})
\begin{align}\label{eq:Htransf}
     H = \sum_{n=1}^N \left[ \frac{p_n^2}{2m} - g \mu_B B \begin{pmatrix} \cos(2 m \alpha_R z_n) \\ - \sin(2 m \alpha_R z_n) \\ 0 \end{pmatrix} \cdot \vec{S}_n\right] + \sum_{n<m}^N V(z_m - z_n).
\end{align}
For $B=0$, the low-density limit of this Hamiltonian has been studied in Refs.~[\onlinecite{matveev04,matveev07}]. Strong repulsions favor a crystalline alignment of the electrons near lattice position $z_n \approx a n$, where $a = 1/\rho$ is the lattice spacing. Including the kinetic energy allows fluctuations about these lattice positions, and gives rise to a single branch of acoustic phonons with wave vector $k \in [-\pi \rho, \pi \rho]$. The charge sector of the system can be described by the Hamiltonian,
\begin{align}\label{eq:Hc}
    H_c = \sum_k \omega(k) a_k^\dag a_k,
\end{align}
where $a_k$ are bosonic operators. For unscreened Coulomb repulsion, the phonon dispersion $\omega(k)$ has a logarithmic singularity at $k = 0$. If screening by a metallic gate at a distance $d$ from the wire is taken into account, the phonon spectrum near $k = 0$ becomes linear, $\omega(k) \propto v_c k$ with a sound velocity $v_c = [2 e^2 \rho \log(8 \rho d)/(\epsilon m)]^{1/2}$.\cite{glazman92} Denoting by $v_F = \pi \rho/(2m)$ the Fermi velocity of the noninteracting electron system, the low-energy continuum limit of Eq.~(\ref{eq:Hc}) is a LL with Luttinger parameter $K_c = v_F/v_c \ll 1$.

To lowest order, the Coulomb repulsion does not affect the spin sector, thus leaving a $2^N$-fold spin degeneracy. The latter is lifted, however, by virtual spin exchange between neighboring lattice sites. Taking this into account, one finds that in the absence of magnetic field, the spins are described by an antiferromagnetic XXX Heisenberg chain,\cite{matveev04} in accordance with the Lieb-Mattis theorem.\cite{lieb62} Including the magnetic field, we obtain the spin Hamiltonian,
\begin{align}\label{eq:Hs}
    H_s = J(\rho) \sum_{n=1}^{N-1} \vec{S}_n \cdot  \vec{S}_{n+1} - g \mu_B \sum_{n=1}^N \vec{B}_n \cdot \vec{S}_n.
\end{align}
with exchange constant $J(\rho) \approx E_F \exp(- \eta/\sqrt{\rho a_B}) \ll E_F$ and $\eta \approx 2.8$.\cite{hausler95,meyer09} In addition to the antiferromagnetic exchange term, the Hamiltonian contains a spiral magnetic field $\vec{B}_n = B [ \cos(\varphi n), \sin(\varphi n), 0]$, where $\varphi$ is defined in Eq.~(\ref{eq:BcritNI}) and symmetry allows us to restrict our analysis to $\varphi \in [0, \pi]$. For $B=0$, the spectrum is gapless and a low-energy limit leads back to a LL Hamiltonian for the spin sector at temperatures $T \ll J$.\cite{fiete07} In contrast, the Wigner crystal remains stable up to much higher temperatures $J \ll T \lessapprox E_F$. We note that due to the dependence of $J$ on $\rho$, the Wigner crystal picture naturally gives rise to the spin-charge coupling expected when going beyond the linear-spectrum approximation of Luttinger theory.\cite{schmidt09_3,imambekov12,schmidt10_2}

The helical gap shows up as an opening of the spectral gap in the spin Hamiltonian (\ref{eq:Hs}) above a critical magnetic field. Before discussing the phase diagram of the Hamiltonian $H_s$, let us discuss some simple limits. On the one hand, for large densities ($\varphi \ll 1$) the magnetic field is essentially constant. In that case, $H_s$ describes a Heisenberg XXZ model, whose phase diagram is well known: the system remains gapless up to a critical magnetic field $\Bcrit = 2J(\rho)$. For larger fields, a gap opens and the spins order ferromagnetically along the applied field.

On the other hand, for $\varphi = \pi$, corresponding to the critical density $\rho = (\pi \lso)^{-1}$, the magnetic field is precisely staggered: $B_n^x = (-1)^n B$. This type of Heisenberg model was investigated using bosonization, and it was found that it is quantum-critical. For $B = 0$, the spectrum is gapless, whereas a finite $B$ opens a gap of order $\Delta/J \propto (B/J)^{2/3}$.\cite{affleck99} Hence, at the critical density, an infinitesimal field is sufficient to open the helical gap.

\begin{figure}[t]
  \centering
  \includegraphics[width=\columnwidth]{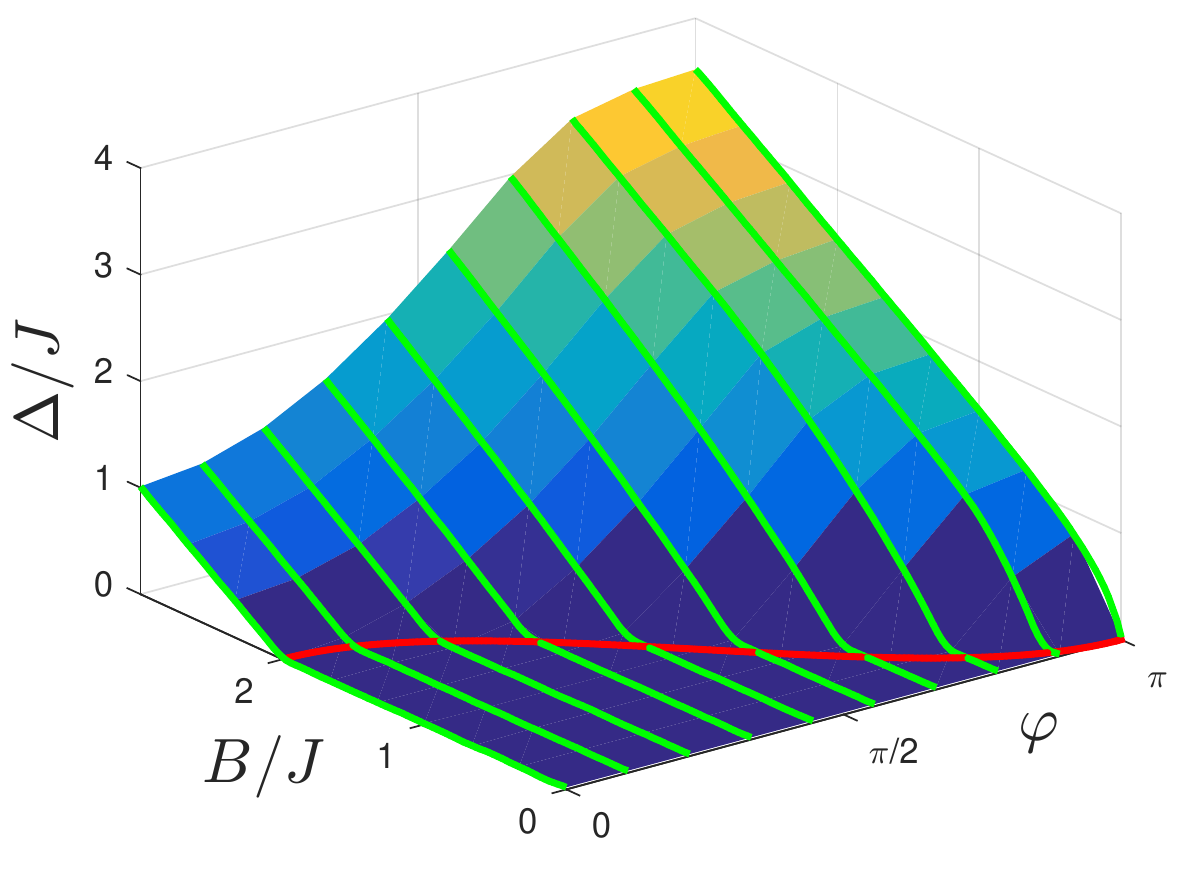}
  \caption{Helical gap $\Delta$ as function of magnetic field $B$ and inverse electron density $\varphi = 1/(\rho\lso)$. The surface plot and the green (bright) lines denote the numerical results obtained using DMRG. The red line shows the critical magnetic field $\Bcrit(\rho)$, see Eq.~(\ref{eq:Bcrit}).}
  \label{fig:HelicalGap}
\end{figure}

\begin{figure}[t]
  \centering
  \includegraphics[width=\columnwidth]{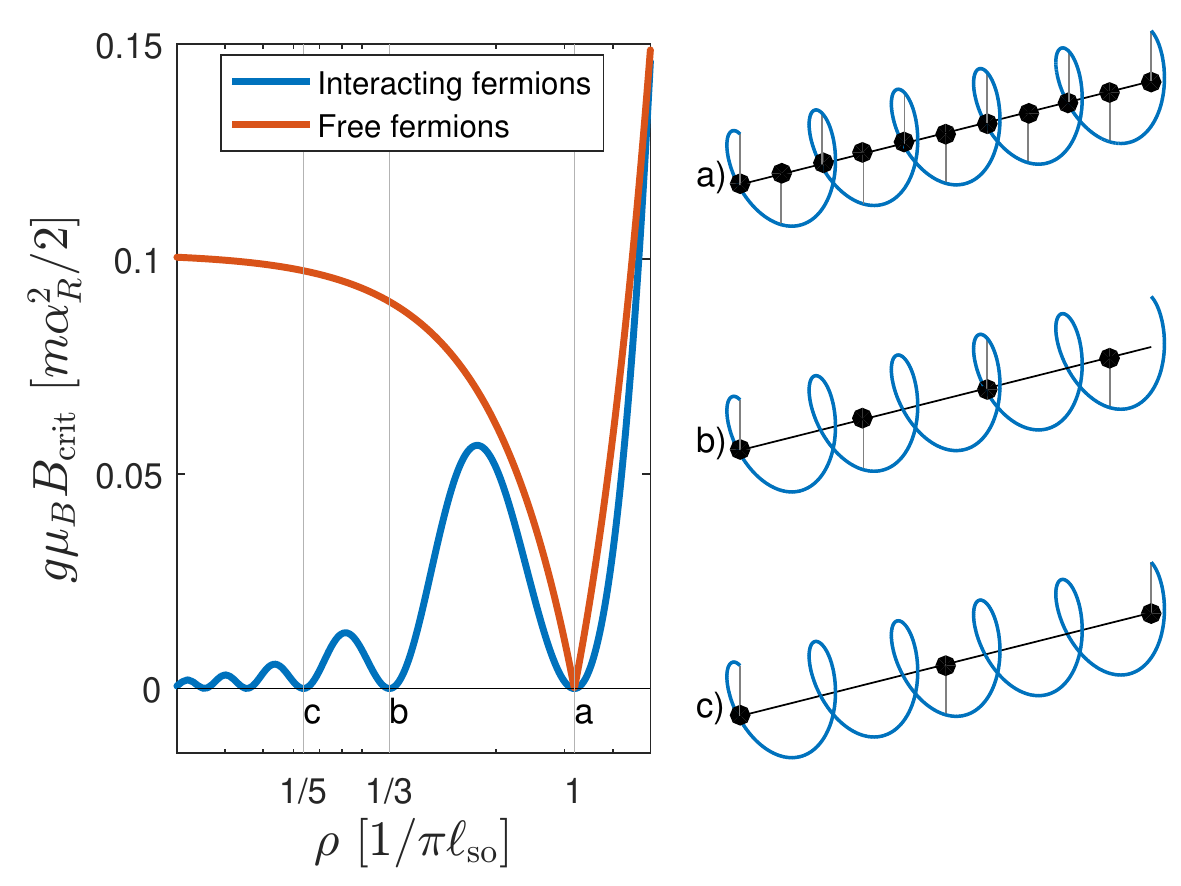}
  \caption{Critical magnetic field $\Bcrit$ as a function of density $\rho$ for the noninteracting (red line) and the interacting (blue line) case. In the noninteracting case, $\Bcrit = 0$ only at the critical density $\pi \rho = 1/\lso$. In the interacting case, in contrast, we find $\Bcrit = 0$ whenever the particle density is commensurate with the pitch of the effective spiral magnetic field. Examples for commensurate densities are shown in the right panel, where the dots denote the electron positions and the spiral indicates the effective magnetic field, see Eq.~(\ref{eq:Htransf}).}
  \label{fig:Bcrit}
\end{figure}

To investigate the full crossover between the limits of constant ($\varphi = 0$) and staggered ($\varphi = \pi$) magnetic fields, we solve the Hamiltonian~(\ref{eq:Hs}) numerically via a density-matrix renormalization group (DMRG) analysis using the ALPS package.\cite{bauer11,dolfi14} The results for the spectral gap as a function of magnetic field for different values of $\varphi$ are shown in Fig.~\ref{fig:HelicalGap}. Comparing different system lengths (from $N = 64$ up to $N=256$) to mitigate finite-size effects, we find by fitting the numerical results that the critical magnetic field as a function of electron density reads
\begin{align}\label{eq:Bcrit}
    g \mu_B \Bcrit(\rho) = J(\rho) \left[\cos\left(\varphi\right) + 1\right], \qquad
    \varphi = \frac{1}{\rho \lso}.
\end{align}
This equation is the central result of this article. It predicts that the critical magnetic field to open a helical gap at a given electron density is actually an oscillatory function of density, in stark contrast to the noninteracting result (\ref{eq:BcritNI}) and results based on Luttinger liquid theory. A comparison between interacting and noninteracting results is shown in Fig.~\ref{fig:Bcrit}. The figure also illustrates that the helical gap can be regarded as a commensurability effect between the pitch of the effective spiral magnetic field and the density. We will now discuss the implications of this result and compare it to existing results.

The expression (\ref{eq:Bcrit}) for the critical field can be reproduced using spin-wave theory. Despite being a large-$S$ expansion, this semiclassical approximation is known to often yield qualitatively correct results even for $S=1/2$ Heisenberg chains.\cite{altland_simons_book} As we show in App.~\ref{sec:SpinWave}, in addition to Eq.~(\ref{eq:Bcrit}), spin-wave theory predicts the following scaling of the helical gap for $B = \Bcrit + \delta B$,
\begin{align}
    \Delta(\rho,B) = g \mu_B \sqrt{\delta B} \sqrt{\delta B + J [1 - \cos(\varphi)]}
\end{align}
Hence, we find the expected linear gap opening $\Delta \propto \delta B$ for $\varphi \ll 1$, similarly to the noninteracting limit in Eq.~(\ref{eq:Delta}). On the other hand, for $\delta B \ll J [1 - \cos(\varphi)]$, spin-wave theory predicts that the gap opens with a square-root cusp, $\Delta \propto \delta B^\gamma$ with $\gamma = 1/2$. The fact that $\Delta(\delta B)$ changes from linear to power-law behavior as $\varphi$ is increased agrees well with our DMRG results. However, the true exponent of the power-law differs from the spin-wave theory prediction. Indeed, from our numerical simulation we find $\gamma \approx 0.66$ at $\varphi = \pi$, in agreement with bosonization result $\Delta \propto \delta B^{2/3}$ for the Heisenberg chain in a staggered magnetic field.\cite{affleck99}

\section{Differential conductance}

\begin{figure}[t]
  \centering
  \includegraphics[width=\columnwidth]{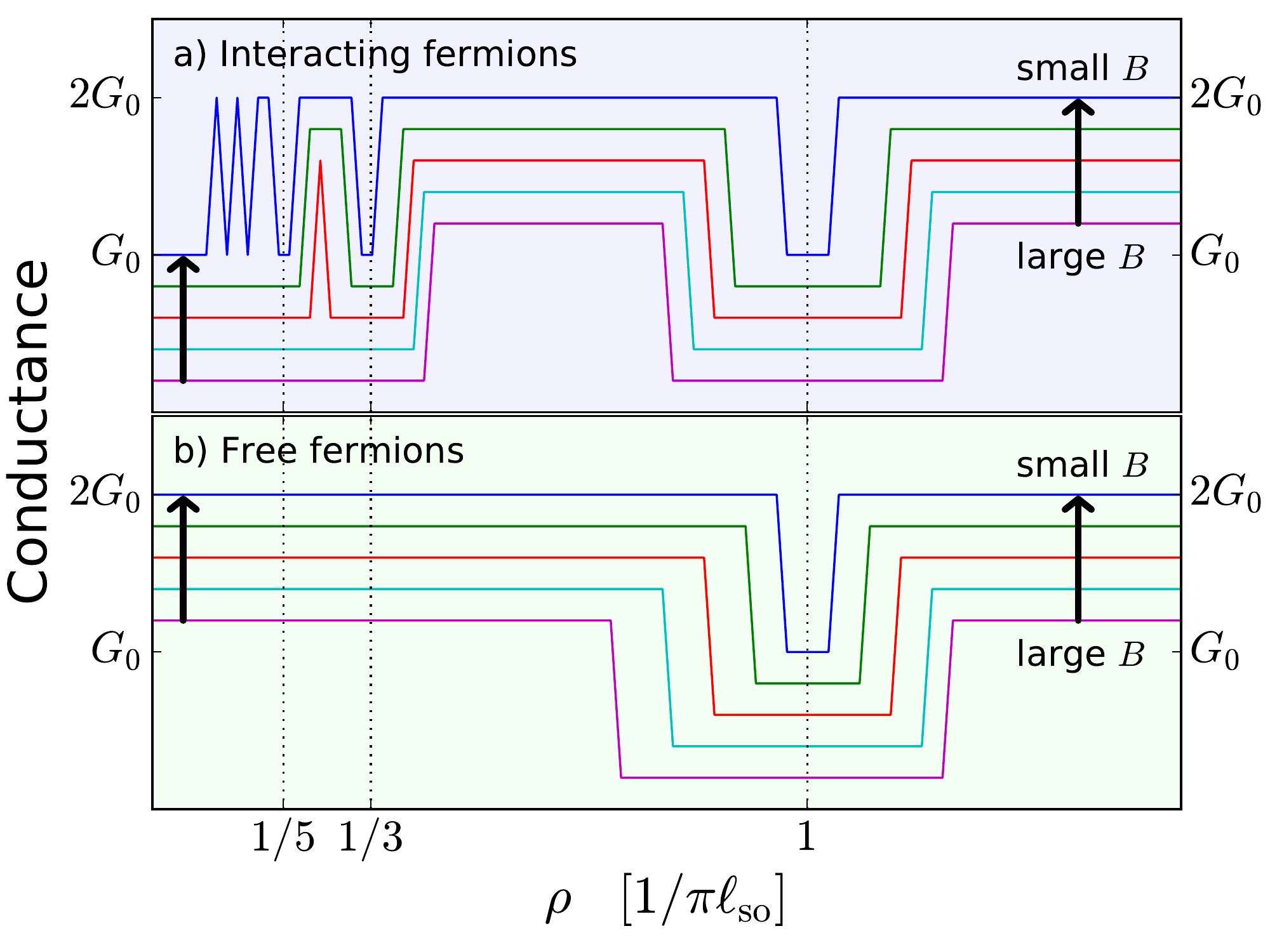}
  \caption{Schematic plots of the conductance $G(\rho)$ for different values of $B$. For clarity, the lines for larger magnetic fields have been shifted downwards. In the interacting case (upper panel), the conductance drops whenever the Wigner lattice is commensurate with the Rashba length. Moreover, the conductance saturates at $G_0$ towards low densities because $J(\rho) \to 0$, in stark contrast to the noninteracting case (lower panel).}
  \label{fig:Conductance}
\end{figure}

A possible way to observe the helical gap which is currently being explored in experiments\cite{vanweperen15,kammhuber16,zhang16} is to study the zero-bias conductance of Rashba wires as the electron density is lowered. Therefore, let us briefly discuss this quantity in the low-density regime. Calculating the conductance of an interacting quantum wire is a nontrivial problem because of the importance of the contacts.\cite{safi95,ponomarenko95,maslov95} In case of Rashba wires, it is known in particular that the contact profile can modify the amplitude of the conductance step.\cite{rainis14} In the Wigner crystal regime, the conductance of a wire with noninteracting contacts can be derived by studying the dissipated heat when the system is subject to an ac drive current $I(t) = I_0 \cos(\omega t)$,\cite{matveev04,matveev04_2,matveev07} and taking the limit $\omega \to 0$. The method was reviewed in detail in Ref.~[\onlinecite{meyer09}]. Adapting it to our system, we find that at low temperatures $T \ll J, \Delta$, the conductance is given by Eq.~(\ref{eq:G}) with the modified critical field $\Bcrit(\rho)$ in Eq.~(\ref{eq:Bcrit}), which is now an oscillatory function of $\rho$. Hence, at the critical density $\rho = (\pi\lso)^{-1}$, the conductance reaches the value $G_0$ and increases towards $2G_0$ in its vicinity. However, as shown in Fig.~\ref{fig:Bcrit} a reduced conductance $G_0$ is reached again at lower densities $\rho \ll (\pi\lso)^{-1}$, whenever the electron density is commensurate with the spin-orbit length. A schematic plot of the conductance as a function of density is shown in Fig.~\ref{fig:Conductance}. For low densities (at fixed $a_B$), $\Bcrit \to 0$, so the conductance is reduced to $G = G_0$ for any finite magnetic field. This is in stark contrast to the behavior for noninteracting or weakly interacting systems [see Eq.~(\ref{eq:BcritNI})], where $g \mu_B \Bcrit = m \alpha_R^2/2$ for $\rho \to 0$, so $G(\rho \to 0, B) = 2 G_0$ for weak magnetic fields $B < \Bcrit$.

Disorder is always an important concern in one dimension, both from the point of view of Luttinger liquids where it is renormalization-group relevant for repulsive interactions, and in the Wigner crystal where it can drive a Peierls instability. In this respect, it is encouraging to note that recent experiments have managed to realize good contacts,\cite{kammhuber16} suspended wires,\cite{montemurro15} and ballistic transport with a mean free path of several $\mu$m in InSb nanowires.\cite{zhang16} We therefore expect our predictions to be observable in these state-of-the-art wires.

\section{Conclusions}

To conclude, we have shown that at the low electron densities $\rho$ needed to see the helical gap in experiments on Rashba wires, Coulomb repulsion dominates over the kinetic energy of electrons. To access this regime, we developed a Wigner crystal theory for 1D systems with RSOC. Within this theory, the helical gap arises in the spin sector as a consequence of commensurability between the Wigner lattice spacing $1/\rho$ and the Rashba length $\lso$. We studied the critical magnetic field for the opening of a helical gap as a function of the electron density. We found that, in contrast to the noninteracting or weakly interacting cases, the critical field is an oscillatory function of density, with strong implications for conductance measurements. Hence, the effect of strong Coulomb interactions need to be taken into account when looking for experimental signatures of the helical gap in Rashba wires.

\begin{acknowledgments}
We would like to thank Tobias Meng and Thomas Sch\"apers for helpful discussions. We acknowledge support by the National Research Fund, Luxembourg under grant ATTRACT 7556175.
\end{acknowledgments}

\appendix
\section{Hamiltonian with Rashba SOC}\label{sec:Ham}

\subsection{Background}

Let us briefly discuss the theoretical approaches taken so far to describe Rashba wires with interactions.

In the case of weak interactions, it is possible to start with the single-particle spectrum of the free fermions (see Fig.~1 of the main text), linearize it near the Fermi points, and use bosonization to account for the interactions.\cite{braunecker09,braunecker09b,braunecker10,gangadharaiah11,stoudenmire11,gambetta15} Without magnetic field, this results in Luttinger Hamiltonians describing the charge and spin sectors. In this language, a magnetic field generates a sine-Gordon term, and a perturbative renormalization group (RG) analysis allows an estimate of the helical gap, which was shown to open for arbitrary repulsive interactions, and increases in magnitude for stronger interactions.

Parts of this approach continue to work for strong interactions (or low densities). Without magnetic field, a linear Rashba term $\propto \alpha_R p \sigma^z$ can simply be gauged away and the Hamiltonian becomes identical to that of SU$(2)$ invariant spinful fermions. In that case, the spectrum in both charge sector and spin sector remains gapless for arbitrarily strong repulsive interactions. The charge sector is then a Luttinger liquid with Luttinger parameter $K_c \ll 1$ and sound velocity $v_c \gg v_F$, $v_F$ being the Fermi velocity of the noninteracting particles. The parameters of the spin sector are $v_s \ll v_F$ and $K_s = 1$ due to SU$(2)$ invariance.\cite{giamarchi03} Luttinger theory in the two sectors remains valid up to energies $E_{c,s} \sim v_{c,s} \rho$, where $\rho$ is the electron density.\cite{matveev07} A magnetic field can again be added to the system as a perturbation. One finds again a helical gap, but this approach is limited to small magnetic fields $g \mu_B B \ll E_s \ll E_c$.

Hence, in the limit of strong interactions, the Luttinger liquid approach suffers from certain shortcomings. Firstly, linearizing the free particle spectrum is not a good starting point for strong interactions. Secondly, the energy range accessible to LL theory tends to zero for strong interactions,\cite{matveev07} a circumstance which has been discussed in detail in the context of spin-incoherent LLs.\cite{fiete07} In our strongly interacting system this means that while LL theory remains correct in the limit of zero energies, it cannot give correct predictions at the interesting magnetic field strengths. Finally, the RG arguments leading to the gap scaling equations are perturbative and valid only for small magnetic fields.

\subsection{Wigner crystal theory}

To extend the existing approaches for the conductance of Rashba wires \cite{gangadharaiah11,stoudenmire11,schmidt13b} towards strong interactions, we build on successful efforts over the past decades to develop a consistent theory for interacting electrons in one dimension at low densities, which is referred to as a 1D Wigner crystal theory.\cite{wigner34,glazman92,schulz93,matveev04,matveev04_2,meyer09} Clearly, the concept of a Wigner crystal in one dimension has to be taken with a pinch of salt, because the Mermin-Wagner theorem rules out a spontaneous breaking of the translation symmetry in the thermodynamic limit. Moreover, long-range crystalline order is evidently unstable against quantum fluctuations in 1D. These issues have been discussed in detail in a recent review.\cite{meyer09}

In the limit of low energies, Wigner crystal predictions agree with LL theory.\cite{giamarchi03} In particular long-range correlations decay as power-laws with interaction-dependent exponents.\cite{schulz93} However, its advantages over LL theory are twofold: firstly, at low densities its energy range of validity is much larger than that of LL theory,\cite{matveev07} and secondly, it often allows a quantitative estimate of the parameters,\cite{matveev04_2} whereas the parameters entering the LL Hamiltonian, i.e., the Luttinger parameter and the sound velocity, are usually phenomenological at strong interactions.\cite{giamarchi03}

We consider the following Hamiltonian which describes $N$ spinful electrons with quadratic spectrum subject to Rashba spin-orbit coupling, a magnetic field perpendicular to the spin-orbit axis, and interactions,
\begin{align}
    H = \sum_{n=1}^N \left[ \frac{p_n^2}{2m} - \alpha_R p_n \sigma^z_n \right] + \frac{1}{2} \sum_{m \neq n} V(z_m - z_n) - \frac{g \mu_B B}{2} \sum_{n =1}^{N} \sigma^x_n
\end{align}
Here, $p_n$ and $z_n$ are the momentum and position operators of the $n$th particle, $\alpha_R$ denotes the strength of the spin-orbit coupling, and $\sigma^{x,z}_n$ denotes Pauli matrices corresponding to the particle $n$. First, it is convenient to remove the spin-orbit coupling by a unitary transformation. Using the action of the translation operator $e^{i z p_0} p e^{-i z p_0} = p - p_0$, we shift the momentum of particle $n$ by $m \alpha_R \sigma^z_n$,
\begin{align}
    U = \exp\left\{ i \sum_n m \alpha_R z_n \sigma^z_n \right\}
\end{align}
Under this transformation, the Hamiltonian becomes
\begin{align}\label{eq:tH}
     \tH &= U H U^\dag \notag \\
&= \sum_{n=1}^N \frac{p_n^2}{2m} + \frac{1}{2} \sum_{m \neq n} V(z_m - z_n)      - N \eso \\
&- \frac{g \mu_B B}{2} \sum_{n =1}^{N} \left[ \cos(2 m \alpha_R z_n) \sigma^x_n - \sin(2 m \alpha_R z_n) \sigma^y_n \right]\notag
\end{align}
where $\eso = m \alpha_R^2/2$.

\subsection{Charge Hamiltonian}

Let us first review the case $B = 0$. In that case, the transformed Hamiltonian $\tH_c := \tH(B=0)$ is independent of Rashba spin-orbit coupling, so we shall just reproduce the known results for a Wigner lattice here.\cite{matveev04,*matveev04_2,*matveev07} If the electron density is sufficiently small, the potential energy will dominate over the kinetic energy term. A Wigner lattice will then form,\cite{wigner34,schulz93} where the electrons are localized approximately at positions $z_n \approx n L/N = a n$. Here, $L$ denotes the length of the system and $a = 1/\rho = L/N$ is the lattice spacing. In that limit, we can introduce the small displacement operator,
\begin{align}\label{eq:shift}
    \oy_n := z_n - a n \ll a
\end{align}
which is canonically conjugate to $p_n$. Expanding to the second order in the displacement, we find
\begin{align}\label{eq:H_harm}
    \tH_c
&\approx \sum_{n=1}^N \frac{p_n^2}{2m}  + \frac{1}{4} \sum_{n=1}^N \sum_{j \neq 0} V''(aj) (\oy_{n+j} - \oy_n)^2 - N \eso
\end{align}
where we used $V'(am - an) = 0$ which holds because the equilibrium positions of the particles minimize the potential energy. The summation over $j$ is over $N-1$ values. We can assume periodic boundary conditions, i.e., $p_n = p_{N+n}$ and $\oy_n = \oy_{n+N}$ to simplify that sum to $\sum_{j=1}^{N-1}$. The Hamiltonian can now easily be diagonalized by Fourier transformation. We introduce the normal modes
\begin{align}
    \Pi_k = \frac{1}{\sqrt{N}} \sum_{n = 1}^{N} e^{-i k a n} p_n, \quad
    Q_k = \frac{1}{\sqrt{N}} \sum_{n = 1}^{N} e^{i k a n} \oy_n
\end{align}
where $k = 2 \pi m /L$ runs over $N$ momenta in the first Brillouin zone. These operators satisfy the canonical commutation relations $[ \Pi_k, Q_{k'} ] = -i \delta_{kk'}$. It is easy to show that this transforms the Hamiltonian to
\begin{align}
    \tH_c
&= \sum_{k} \left[ \frac{\Pi^\dag_k \Pi_{k}}{2m} + \frac{1}{2} m \omega^2(k) Q^\dag_k Q_{k} - \eso \right]
\end{align}
where we introduced the mode frequencies
\begin{align}
    \omega^2(k)
&=
    \frac{1}{m} \sum_{j=1}^{N-1} V''(aj) [ 1 - \cos(ka j)]
\end{align}
If we were to consider a short-range potential, we would only keep the terms $j = 1$ and $j = N-1$, in which case we would find the typical spectrum of acoustic phonons,
\begin{align}
    \omega(k)
&=  2 \sqrt{V''(a)/m} |\sin(ak/2)|
\end{align}
which corresponds to a linear spectrum for small $k$. On the other hand, for a generic interaction potential, we should express $\omega(k)$ in terms of the Fourier transform of the interaction potential. Using $V(x) = (1/N) \sum_q e^{i q x} V_q$, we have
\begin{align}
    \omega^2(k)
&=
    \frac{k^2}{m} V_k
\end{align}
Finally, we introduce the conventional creation and annihilation operators,
\begin{align}
    \Pi_k &= i \sqrt{\frac{m \omega(k)}{2}} \left( a_k^\dag - a_{-k} \right) \notag \\
    Q_k &= \frac{1}{\sqrt{2 m \omega(k)}} \left( a_{-k}^\dag + a_k \right)
\end{align}
which leads to the Hamiltonian
\begin{align}\label{eq:tHharmonic}
    \tH_c
= \sum_{k} \omega(k) \left( a_k^\dag a_k - \frac{1}{2} \right)  - N \eso
\end{align}
which coincides, up to a constant, with Eq.~(8) in the main text. Because $\omega(k) \to 0$ for $k \to 0$, the excitation spectrum is gapless (except for a trivial finite size gap $\propto 1/L$). The low-energy excitations are acoustic phonons with spectrum $\omega(k) \propto k$. Its eigenstates are Fock states with a certain set of phonon quantum numbers. Each of the eigenvalues has a degeneracy $2^N$ because the eigenenergies are spin-independent within our approximation. A complete basis of this Hamiltonian is given by the vectors
\begin{align}
    \left\{ \ket{n_{k_1}, \ldots, n_{k_N}, \sigma^z_1, \ldots \sigma^z_N} \right\}
\end{align}
where $n_k \in \mathbb{N}_0$ denotes the number of phonons in mode $k$ and $\sigma^z_n \in \{-1, 1\}$ denotes the $z$ component of the spin on lattice site $n$.

\subsection{Spin exchange}

The spin degeneracy is due to the fact that we assumed $B = 0$ and restricted the position of each electron to one site in the Wigner lattice. The most important process we neglected so far is tunneling between neighboring sites. Due to the strong interactions, each lattice site should always be singly occupied. But even in this limit spin exchange between neighboring sites is possible, albeit weak. In order to investigate this effect, we follow Ref.~[\onlinecite{matveev04}] and consider the positions of $N-2$ particles as fixed, and only investigate the dynamics of the two remaining particles.

Starting from Eq.~(\ref{eq:tH}) and keeping $B=0$, these assumptions lead to the two-particle Hamiltonian,
\begin{align}
    \tH_2 = \frac{p_1^2}{2m} + \frac{p_2^2}{2m}  + V(z_1 - z_2) + V_r(z_1) + V_r(z_2)
\end{align}
where $V_r(x)$ denotes the potential generated by the remaining $N-2$ stationary electrons. The two particles are in a double-well potential, which we shall call $U(z)$. Such a scenario was investigated in Ref.~[\onlinecite{matveev04}] for an unscreened Coulomb potential $V(z) = e^2/(\epsilon |z|)$, and a formula for $U(z)$ was derived there. The effective Hamiltonian in real space is now a two-body problem,
\begin{align}
    \left[ - \frac{\partial_{z_1}^2}{2m}  - \frac{\partial_{z_2}^2}{2m}  + U(z_1 - z_2) \right] \phi(z_1,z_2) = E \phi(z_1, z_2)
\end{align}
It is known that the ground state wave function $\phi_S$ is symmetric in $z_1$ and $z_2$, whereas the first excited state $\phi_A$ is antisymmetric.\cite{matveev04} The two states are split by an energy which can be determined using the WKB approximation
\begin{align}
    J = \frac{U''(a)}{m \sqrt{e \pi}} \exp \left\{ - \int_{-z_0}^{z_0} dz \sqrt{2 m \left[ U(z) - U''(a)/2m \right]} \right\}
\end{align}
where $\pm z_0$ are the edges of the classically forbidden region of the potential $U(z)$. Importantly, this energy splitting is independent of the spins of the two particles. Therefore, we can construct the following ground state and first excited state wavefunctions, which consist of a spin-independent orbital part, and a singlet or triplet spin part. The ground state wave function is nondegenerate and reads,
\begin{align}
    \psi_0(z_1,\sigma_1,z_2,\sigma_2) = \phi_S(z_1,z_2) \left[ \delta_{\sigma_1\uparrow} \delta_{\sigma_2\downarrow} - \delta_{\sigma_1\downarrow} \delta_{\sigma_2\uparrow} \right]
\end{align}
The first excited state is a threefold degenerate triplet and reads,
\begin{align}
    \psi_{1,-1}(z_1,\sigma_1,z_2,\sigma_2) &=  \phi_A(z_1,z_2) \delta_{\sigma_1\downarrow} \delta_{\sigma_2\downarrow} \notag \\
    \psi_{1,0}(z_1,\sigma_1,z_2,\sigma_2) &= \phi_A(z_1,z_2) \left[ \delta_{\sigma_1\uparrow} \delta_{\sigma_2\downarrow} + \delta_{\sigma_1\downarrow} \delta_{\sigma_2\uparrow} \right] \notag \\
    \psi_{1,1}(z_1,\sigma_1,z_2,\sigma_2) &=  \phi_A(z_1,z_2) \delta_{\sigma_1\uparrow} \delta_{\sigma_2\uparrow}
\end{align}
If we are only interested in the spin degrees of freedom, we can therefore describe this by a Hamiltonian,
\begin{align}
    \tH_2 = J \vS_1 \cdot \vS_2
\end{align}
The alignment of nearest neighbors' spins is antiferromagnetic in accordance with the Lieb-Mattis theorem.\cite{lieb62} So far, we showed this for two sites. But since next-nearest-neighbor hopping is exponentially suppressed compared to nearest-neighbor hopping, we can use the following Heisenberg Hamiltonian for the spin system
\begin{align}
    \tH_s = J \sum_{n = 1}^{N} \vS_n \cdot \vS_{n+1}
\end{align}
It should be pointed out that $J$ depends on the positions of the electrons and may in principle be nonuniform, $J \to J_n$. In that sense, $\tH_s$ implicitly contains spin-charge coupling. Treating $J$ as a constant works as long as $\oy_n \ll a$.

\section{Spin wave theory}\label{sec:SpinWave}

We start from the Hamiltonian~(9) in the main text. To investigate the case for general $\varphi$, it is convenient to restore translation invariance by mapping the system with spiral magnetic field onto a system with constant magnetic field and modified exchange terms,
\begin{align}\label{eq:Hsprime}
    H'_s
&=
    J(\rho) \sum_{n=1}^{N-1} \Big[ \cos(\varphi) (S^x_n S^x_{n+1} + S^y_n S^y_{n+1})  \\
&+ \sin(\varphi) (S^x_n S^y_{n+1} - S^y_n S^x_{n+1}) + S^z_n S^z_{n+1} \Big] - g \mu_B  B \sum_{n=1}^N S^x_n.\notag
\end{align}
As we are mainly interested in physical effects near the critical field, we use the ferromagnetic large-field state as a starting point for spin-wave theory. For $B \gg J > 0$, the spins are all polarized in the $+x$ direction, and we can use the Holstein-Primakoff transformation where the largest component is in $x$ direction,
\begin{align}
    \vec{S}_n \approx \left( S - c^\dag_n c_n,\quad \sqrt{\frac{S}{2}} (c^\dag_n + c_n), \quad i\sqrt{\frac{S}{2}} ( c_n^\dag - c_n) \right)
\end{align}
where $c_n$ and $c_n^\dag$ are bosonic annihilation and creation operators and $S = 1/2$. Hence, we obtain the following terms,
\begin{widetext}
\begin{align}
    J \cos(\varphi) S^x_n S^x_{n+1}
&=
    J \cos(\varphi) S^2 - JS \cos(\varphi) c^\dag_n c_n - JS \cos(\varphi) c^\dag_{n+1} c_{n+1} + \text{irrelevant terms} \notag \\
    J \cos(\varphi) S^y_n S^y_{n+1}
&=
    \frac{J \cos(\varphi) S}{2} \left( c^\dag_n c^\dag_{n+1} + c^\dag_n c_{n+1} + c_n c^\dag_{n+1} + c_n c_{n+1} \right) \notag \\
    J S^z_n S^z_{n+1}
&=
    - \frac{J S}{2} \left( c^\dag_n c^\dag_{n+1} - c^\dag_n c_{n+1} - c_n c^\dag_{n+1} + c_n c_{n+1} \right) \notag \\
    J \sin(\varphi) S^x_n S^y_{n+1}
&=
    J \sin(\varphi) \left[ \frac{S\sqrt{S}}{\sqrt{2}}(c^\dag_{n+1} + c_{n+1}) - \sqrt{\frac{S}{2}} c^\dag_n c_n (c^\dag_{n+1} + c_{n+1})
     \right]\notag \\
    J \sin(\varphi) S^y_n S^x_{n+1}
&=
    J \sin(\varphi) \left[ \frac{S\sqrt{S}}{\sqrt{2}}(c^\dag_{n} + c_{n}) - \sqrt{\frac{S}{2}} (c^\dag_{n} + c_{n}) c^\dag_{n+1} c_{n+1}      \right]
\end{align}
\end{widetext}
Spin-wave theory is based on a large-$S$ expansion. When summed over $n$ in the Hamiltonian the terms $\propto S^{3/2}$ cancel. Moreover, the terms $\propto \sqrt{S}$ are subleading and can be ignored. To do a systematic expansion, we assume that $B$ is also of order $S$,\cite{affleck99} and keep only the terms of order $S$. We obtain, after Fourier transform,
\begin{align}
    H
&=
    \sum_{k} \bigg\{
    (- 2 J S \cos(\varphi) + B) c^\dag_k c_k \notag \\
&+
    \frac{J S}{2} \left( \cos(\varphi) - 1\right) \left( e^{ik} c^\dag_k c^\dag_{-k} + \hc \right) \notag \\
&+
    \frac{J S}{2} \left( \cos(\varphi) + 1\right) \left( e^{ik} c^\dag_k c_{k} + \hc \right)
    \bigg\}
\end{align}
Therefore, we can write this as
\begin{align}
    H
&=
    \frac{1}{2} \sum_k \begin{pmatrix} c_k^\dag \\ c_{-k} \end{pmatrix}^T
    \begin{pmatrix}
    X & Y \\ Y^* & X
    \end{pmatrix}
    \begin{pmatrix} c_k \\ c^\dag_{-k} \end{pmatrix}
\end{align}
where
\begin{align}
    X &= - 2 J S \cos(\varphi) + B + JS (\cos(\varphi)  + 1) \cos(k) \notag \\
    Y &= JS (\cos(\varphi) - 1) e^{ik}
\end{align}
We solve the Hamiltonian using a Bogoliubov transformation. We write the operators as $c^\dag_k = u b^\dag_k + v b_{-k}$ and $c^\dag_{-k} = u b^\dag_{-k} + v b_k$. If we assume that $b_k$ fulfill bosonic commutation relations, this leads to $[c_k, c_{k'}^\dag] = (|u|^2 - |v|^2) \delta_{kk'}$ and thus to the requirement $|u|^2 - |v|^2 = 1$, which we can satisfy by setting $u = e^{i \phi_1} \cosh \theta$ and $v = e^{i \phi_2}\sinh \theta$. In terms of the new operators, we find
\begin{align}
    H
&=
    \frac{1}{2} \sum_k \begin{pmatrix} b_k^\dag \\ b_{-k} \end{pmatrix}^T
    \begin{pmatrix}
    X' & Y'  \\
    Y'^*  & X'
    \end{pmatrix}
    \begin{pmatrix} b_k \\ b^\dag_{-k} \end{pmatrix} \notag \\
    X' &= X \cosh(2\theta) + Y \cos(\phi_1 + \phi_2) \sinh(2\theta) \notag \\
    Y' &= e^{i (\phi_1 - \phi_2)} \Big[ X \sinh(2 \theta) \notag \\
    & + Y e^{i (\phi_1 +\phi_2)} \cosh^2(\theta) + Y^* e^{-i(\phi_1+ \phi_2)} \sinh^2(\theta) \Big]
\end{align}
We would like to choose the parameters in such a way that the off-diagonal part vanishes. We can achieve this by first demanding that $Y e^{i (\phi_1 +\phi_2)}$ is real, i.e., $\phi_1 +\phi_2 = - \arg Y$. Having fixed this, vanishing off-diagonal elements leads to
\begin{align}
    \tanh(2\theta) = - \frac{|Y|}{X}
\end{align}
which has a real solution only if $|Y| < X$. With these parameters, the Hamiltonian takes the form $H = \sum_k \epsilon(k, \varphi) b_k^\dag b_k$ with the eigenenergies
\begin{align}
    \epsilon(k,\varphi) := X'
&=
    \frac{X^2 - |Y| \text{Re} Y}{\sqrt{ X^2 - |Y|^2}}
\end{align}
Unfortunately, in general there seems to be no simple solution of the energies. We find, however, rather simple expression for $\varphi = 0$ and $\varphi = \pi$,
\begin{align}
    \epsilon(k,\varphi = 0) &= \left| J [\cos (k)-1] + B\right| \notag \\
    \epsilon(k,\varphi = \pi) &= \frac{(B+J)^2+J^2 \cos (k)}{\sqrt{B (B+2 J)}}
\end{align}
We see that $\epsilon(k,\varphi =0)$ is gapped for all $B > 2J$. In contrast, $\epsilon(k,\varphi = \pi)$ is gapped for all $B > 0$. These limits coincide with Eq.~(10) of the main text. To study the behavior for arbitrary $\varphi$, we observe by plotting the general function $\epsilon(k,\varphi)$ that when increasing $J$ for fixed $B$, the gap closing always occurs at $k = \pi$. In this case, we also find a simple result
\begin{align}
    \epsilon(k = \pi,\varphi) = \sqrt{(B- 2 J \cos \varphi) (B-J-J \cos \varphi)}
\end{align}
From this equation, one finds Eq.~(10) of the main text as the condition for the having a finite gap. Expanding $\epsilon(k=\pi,\varphi)$ close to this critical field, i.e., using $B = J (1 + \cos \varphi) + \delta B$, we find
\begin{align}
    \epsilon(k = \pi,\varphi,\Bcrit + \delta B) = \sqrt{ (J - J \cos \varphi) + \delta B ) \delta B}
\end{align}
For $\varphi = 0$, the gap opens indeed linearly, $\epsilon \propto \delta B$. On the other hand, for $J (1 - \cos \varphi) \gg \delta B$, the gap opens with a square root dependence.

\section{Connection to bosonization}\label{sec:Bosonization}

The helical gap for interacting systems was studied in Ref.~[\onlinecite{stoudenmire11}] based on Luttinger theory, and we would like to connect to their results. Such a comparison is possible exactly at the critical density, i.e., at chemical potential $\mu = 0$, where the chemical potential is exactly at the band crossing. In this case, the effective magnetic field after removing the Rashba spin-orbit coupling is just staggered, the critical field for opening a gap vanishes, and we can compare the gap width as a function of magnetic field.

To use bosonization for chemical potential $\mu = 0$, we linearize the spectrum around $k = \pm k_F = \pm 2 m \alpha_R$ and $k = 0$ and introduce left-moving and right-moving fermionic operators by decomposing the physical fermions as,\cite{stoudenmire11}
\begin{align}
    \psi_\uparrow &= \psi_{L\uparrow} + e^{i k_F x} \psi_{R\uparrow} \notag \\
    \psi_\downarrow &= e^{-i k_F x} \psi_{L\downarrow} + \psi_{R\downarrow}
\end{align}
We bosonize these according to $\psi_{\alpha\sigma} = (2 \pi a)^{-1/2} e^{-i ( \alpha \phi_\sigma - \theta_\sigma)}$, where $\alpha = R, L = +,-$, $\sigma = \uparrow,\downarrow$, and $a$ denotes the short-distance cutoff. Next, we introduce charge and spin modes, $\phi_{c,s} = (\phi_\uparrow \pm \phi_\downarrow)/\sqrt{2}$ and analogously for $\theta_{c,s}$. In the absence of magnetic field, the resulting Luttinger Hamiltonian is characterized by two Luttinger parameters, $K_c$ and $K_s$, for the charge and spin sector, respectively. In the limit of strong repulsive interactions, we have $K_c \ll 1$, whereas $K_s = 1$ is fixed by SU(2) symmetry.

Next, we add the magnetic field term, which couples to a linear combination of charge and spin modes,
\begin{align}
    H_B
&=
    B \int dx \left[ \psi^\dag_\uparrow(x) \psi_\downarrow(x) + \hc \right] \notag \\
&=
    \frac{B}{\pi a} \int dx \cos\left[ \sqrt{2} (\phi_c - \theta_s ) \right]
\end{align}
In Ref.~[\onlinecite{stoudenmire11}] it was found using an RG analysis that to leading order the Zeeman term obeys the following scaling equation,
\begin{align}
    \frac{dB}{d\ell} &= \frac{(3 - K_c)}{2} B \notag \\
    B(\ell) &= B(0) e^{\gamma \ell}
\end{align}
where $\gamma = (3 - K_c)/2$. Here, $\ell$ is the logarithmically scaled cutoff, and is related to the physical cutoff as $a(\ell) = a_0 e^{-\ell}$. $H_B$ is thus a relevant perturbation. At the end of the RG flow (say, at $\ell = \ell^*)$, $H_B$ dominates and generates a gap proportional to $B$, hence $\Delta(\ell^*) \approx B(\ell^*)$. From this we can calculate the bare gap,\cite{giamarchi03}
\begin{align}
    \Delta(0) = e^{-\ell^*} \Delta(\ell^*) = \left(\frac{B(0)}{B(\ell^*)}\right)^{1/\gamma} B(\ell^*) \propto B(0)^{1/\gamma}
\end{align}
Therefore, at $\mu = 0$, we find in the limits of weakly interacting and strongly interacting fermions, respectively,
\begin{align}
    \begin{aligned}
    \Delta(B) &\propto B && \text{for } K_c = 1 \\
    \Delta(B) &\propto B^{2/3} && \text{for } K_c = 0.
    \end{aligned}
\end{align}
The exponent $1$ for $K_c = 1$ agrees with the trivial noninteracting result, see Eq.~(4) in the main text. The exponent $2/3$ in the strongly interacting limit agrees with what we found from our Heisenberg model from the DMRG solution at the staggered point and from the bosonization solution of the corresponding Heisenberg model,\cite{affleck99} see Eq.~(11) in the main text, and the following paragraph.

\section{Screening}\label{sec:Screening}

Wigner crystal ordering can be expected when the potential energy per particle dominates over the kinetic energy. Whereas the latter scales as $E_{\rm kin} \propto \rho^2$ due to the Pauli principle, the former scales as $E_{\rm pot} \propto \rho^{1+\gamma}$ if the physical interaction potential between the electrons behaves as $V(z) \propto z^{-1-\gamma}$. Hence, for an unscreened Coulomb potential ($\gamma = 0$), one finds always $E_{\rm pot} > E_{\rm kin}$ for low densities.

\begin{figure}[t]
  \centering
  \includegraphics[width=\columnwidth]{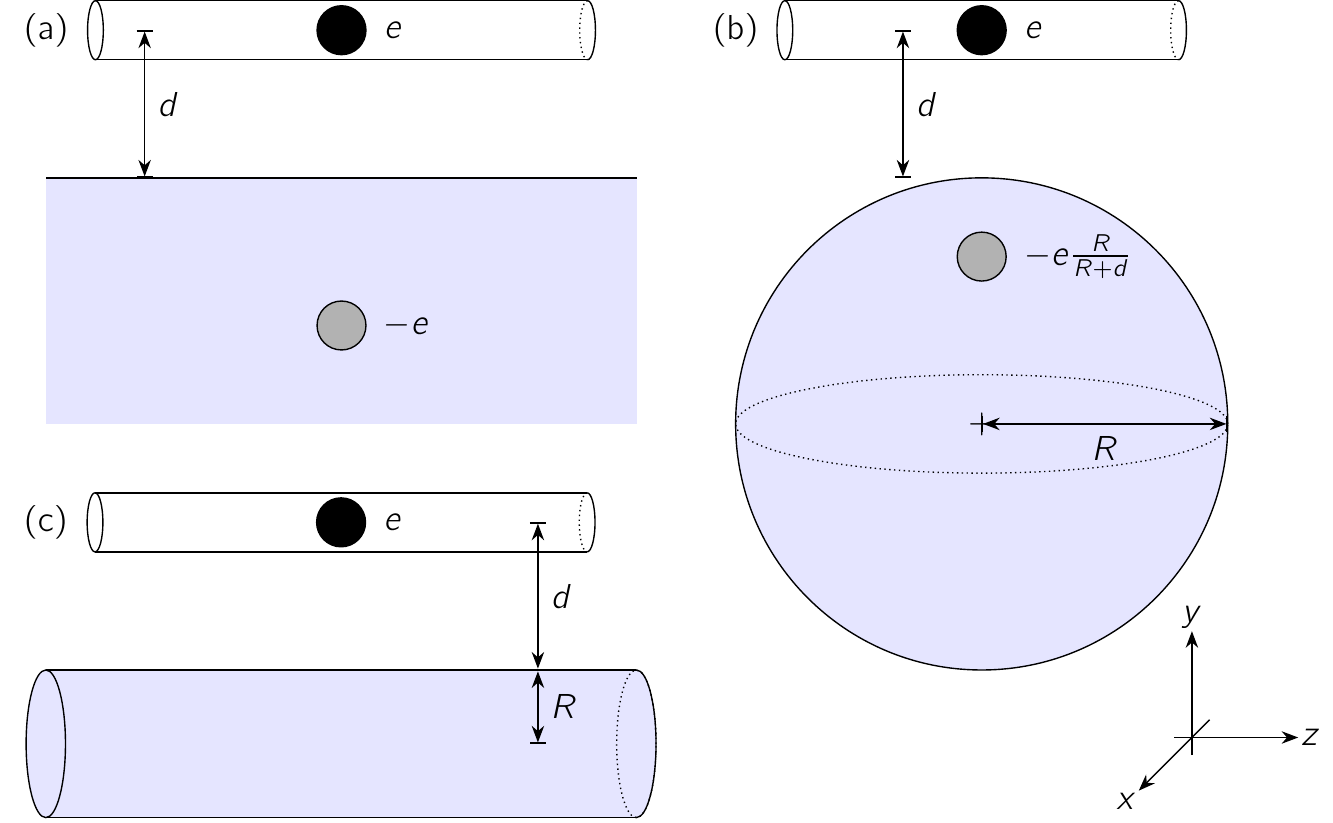}
  \caption{Different models for screening Coulomb interactions due to a metal at distance $d$ from an electron with charge $e$ in a nanowire along the $z$ direction: (a) infinite metallic surface, (b) metallic sphere, (c) metallic cylinder.}
  \label{fig:Screening}
\end{figure}

However, in many realistic situations, the interaction potential decays faster than $1/z$ for large $z$ due to screening ($\gamma > 0$). In that case, the condition $E_{\rm pot} > E_{\rm kin}$ can again be violated for too low densities. Here, we will discuss how this condition is affected by screening due to nearby metallic gates held at fixed electric potential. The screening models we compare are depicted in Fig.~\ref{fig:Screening}.

If the gate is modelled as an infinite metallic surface in the $x-z$ plane, at a distance $d$ to a point charge in the nanowire at position $\vr_0 = (0, d, 0)$, the theory of image charges predicts that the total electric potential at some position $\vr$ is given by
\begin{align}\label{eq:Phi_image}
    \Phi(\vr) = \frac{e}{|\vr - \vr_0|} + \frac{q}{|\vr - \vr_q|}
\end{align}
where $\vr_q = (0, -d, 0)$ is the position of the image charge and its charge $q = -e$. If we place a probe charge in nanowire at position $\vr = (0, d, z)$, the total potential acting on it will be given by
\begin{align}
    \Phi(z) = \frac{e}{|z|} - \frac{e}{\sqrt{z^2 + 4 d^2}} \approx \frac{2d^2}{|z|^3}\ \text{for } z \gg d
\end{align}
Hence, screening by an infinite metallic plane parallel to the nanowire leads to the screened Coulomb potential discussed in the main text. The scaling $\propto z^{-3}$ for $z \to \infty$ corresponds to $\gamma = 2$, so this screened potential gives a \emph{low-density} threshold for the Wigner crystal formation. This model of an infinite surface can be used to describe screening due to nearby macroscopic metallic gates, such as a back gate used to deplete the nanowire.

\begin{figure}[t]
  \centering
  \includegraphics[width=\columnwidth]{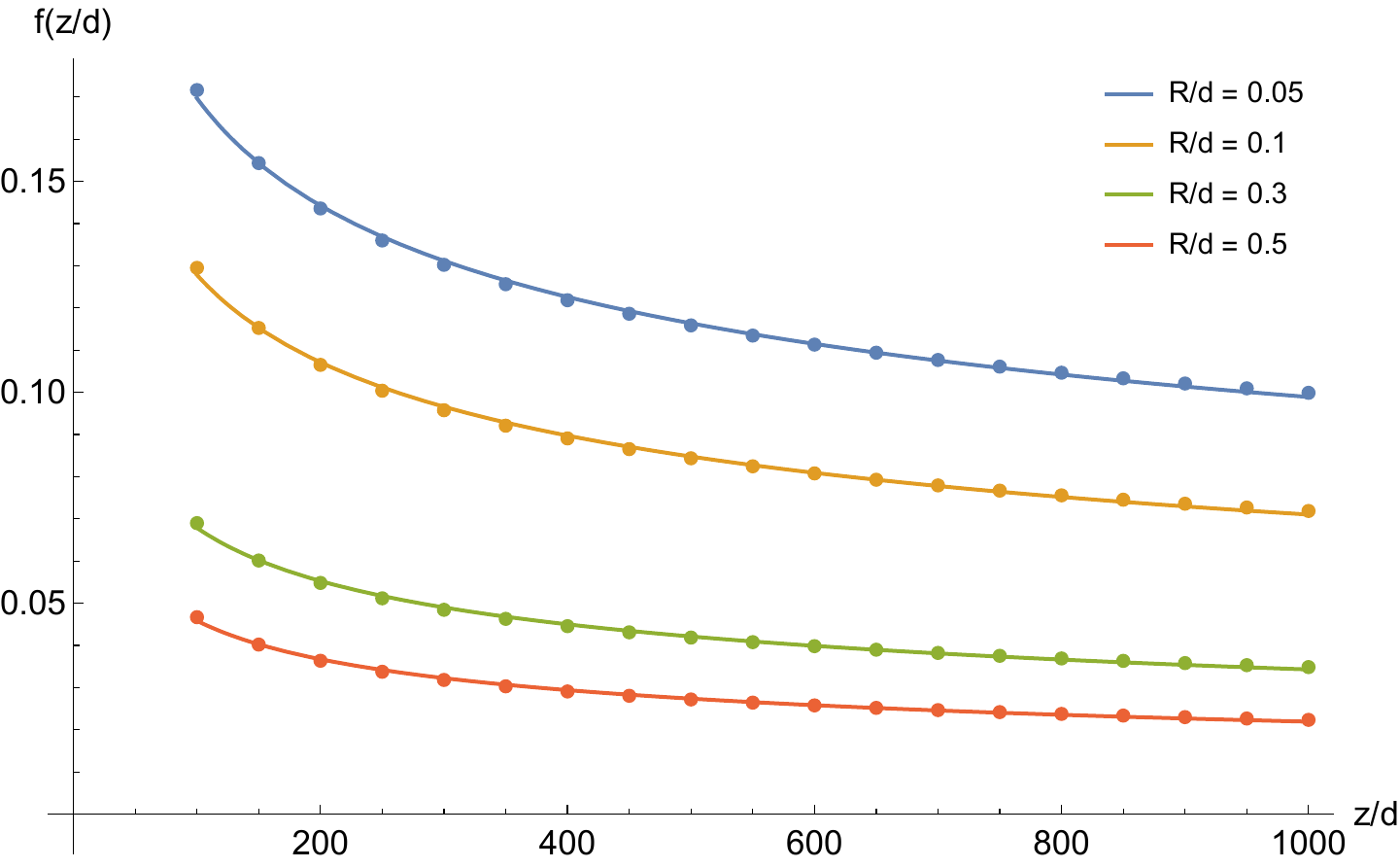}
  \caption{Correction to the Coulomb potential due to screening by a metallic cylindrical gate with radius $R$ at distance $d$ to the wire.}
  \label{fig:ScreeningPlot}
\end{figure}

Let us compare this with screening due to a metallic gate with finite size. A simple example will be screening due to a metallic sphere. We consider a metallic sphere with radius $R$ centered at $(0, -R, 0)$. In this case, it is well known that the total potential of point charge and sphere is again given by Eq.~(\ref{eq:Phi_image}), but with
\begin{align}
    q &= -e \frac{R}{R+d}  \notag \\
    \vr_q &= \left(0, - d \frac{R}{R+d} - 1, 0 \right)
\end{align}
For an infinite sphere ($R \to \infty$), the result is equivalent to the previous one of the metallic surface. For finite $R$, however, the image charge is less that the electron charge, $|q| < |e|$. Therefore, a probe charge far away at position $\vr = (0, d, z)$ experiences the potential,
\begin{align}
    \Phi(z) = \frac{e}{|z|} \frac{d}{d + R} + O\left(\frac{1}{z^3} \right)
\end{align}
Therefore, the $1/z$ behavior remains the leading contribution for all finite $R$. Hence, the finite sphere cannot screen the Coulomb interacting at large distances and the Wigner crystal formation remains possible up to the lowest densities. This model for screening applies, for instance, to gate wires running perpendicular to the nanowire, which as a consequence do not significantly screen the Coulomb repulsion.

In order to create low electron densities in the wires, some form of metallic gate running parallel to the nanowire axis is usually necessary. It can be modelled most realistically as a cylindrical wire parallel to the nanowire. Calculating the potential of a single point charge at a distance $d$ from a metallic cylinder with radius $R$ is a complicated electrostatics problem and was solved only recently using Green's functions.\cite{hernandes05} The resulting equation contains an integral over a product of Bessel functions, which can be evaluated numerically. One finds that the leading long-range asymptotic behavior for $z \gg d$ is given by
\begin{align}
    \Phi(z) = \frac{e}{|z|} f\left(\frac{z}{d}, \frac{R}{d}\right)
\end{align}
where the function $f(z/d)$ is plotted in Fig.~\ref{fig:ScreeningPlot} for different values of $R/d$. This function is approximately given by $f(x) = a x^{-\gamma}$. The prefactor $a$ depends strongly on $R/d$, the exponent is approximately $\gamma \approx 0.3$ for the values of $R/d$ considered. Only towards $R \gg d$ does one find again a stronger correction to the exponent and ultimately one recovers the behavior $\Phi(z) \propto 1/|z|^3$ for $R \to \infty$. This allows us to conclude that for the experimentally realistic distances, i.e., for $d$ on the same order as $R$, a thin cylindrical gate cannot effectively screen a the Coulomb interaction and the long-range behavior gets only weakly modified.

From this discussion, we can conclude that if only thin gate wires are present, the Wigner crystals remains intact up to the lowest densities. Therefore, the low-density threshold for the Wigner crystal formation will be rather given by a macroscopic metallic depletion gate.

\bibliography{references}

\end{document}